\documentclass[fleqn,usenatbib]{mnras}
\usepackage{newtxtext,newtxmath}
\usepackage[T1]{fontenc}
\DeclareRobustCommand{\VAN}[3]{#2}
\let\VANthebibliography\thebibliography
\def\thebibliography{\DeclareRobustCommand{\VAN}[3]{##3}\VANthebibliography}

\newcommand{\ud}{\mathrm{d}}

\usepackage{graphicx}	
\usepackage{amsmath}	
\usepackage{caption}  
\usepackage{subcaption}
\usepackage{tcolorbox}

\title[Synthetic Velocity Field]{3D Synthetic Convective Velocity Fields to Initialise Core-Collapse Supernova Simulations from 1D Progenitors}

\author[Varma, M\"uller \& Hirschi]{
Vishnu Varma$^{1}$\thanks{E-mail: 
v.r.vejayan@keele.ac.uk},
Bernhard M\"uller$^{2}$,
Raphael Hirschi$^{1, 3}$
\\
$^{1}$
{Astrophysics Group, Lennard-Jones Laboratories, Keele University, Keele ST5 5BG, UK}\\
$^{2}$
School of Physics and Astronomy, 10 College Walk, Monash University, Clayton, VIC 3800, Australia\\
$^{3}$
Kavli IPMU (WPI), University of Tokyo, 5-1-5 Kashiwanoha, Kashiwa 277-8583, Japan\\
}

\date{Accepted XXX. Received YYY; in original form ZZZ}

\pubyear{\the\year{}}

\begin{document}
\label{firstpage}
\pagerange{\pageref{firstpage}--\pageref{lastpage}}
\maketitle

\begin{abstract}
Core-collapse supernovae (CCSNe) are among the most energetic and complex astrophysical phenomena, 
requiring three-dimensional (3D) simulations to capture their intricate explosion mechanisms. One of the key ingredients for such simulations is the 3D pre-collapse structure, which can impact the development and geometry of the subsequent explosion.
While stellar convection simulations can provide such 3D initial conditions, these remain too expensive and demanding for widespread use. In this work, we present a method to generate 
synthetic 3D velocity fields for convective zones from 1D initial conditions, creating  initial 
conditions for CCSN simulations using a
vector spherical harmonics expansion
without the need for expensive hydrodynamic progenitor simulations. 
The synthetic velocity field is designed to
capture the typical scales and velocities of the convective flow as the most relevant parameters for the subsequent explosions.
In addition, it respects relevant physical
constraints such as the near-anelasticity of flow, vanishing radial vorticity, and zero net angular momentum in the convective zones.
A \textsc{Python} implementation of this method is publicly available, offering the CCSN community a practical tool 
for generating synthetic velocity fields for multi-dimensional simulations to study the impact of 3D progenitor asymmetries on the CCSN mechanism. 
\end{abstract}
\begin{keywords}
convection -- stars: massive -- supernovae: general 

\end{keywords}



\section{Introduction}
In the decades since the first numerical simulations of core-collapse supernovae (CCSNe) \citep{Colgate1966, Bethe1985},
the field has made significant progress in understanding the neutrino-drive explosion mechanism \citep{Janka2025,Mueller_LRCA,Mezzacappa_LRCA, Burrows2021}.
The increase in computational power and improvements in numerical methods in this time has allowed 
models to include more realistic physics, such as realistic equation of states \citep[e.g.,][]{Suwa2013, Fischer2017}, general relativistic 
gravity \citep{Muller2012}, elaborate methods for neutrino transport and better neutrino opacities \citep[e.g.,][]{Mezzacappa_LRCA, melson_15,lentz_12,horowitz_17,Nagakura2019}, 
and magnetic fields \citep[e.g.,][]{Winteler2012,Obergaulinger2014,Moesta_2014, mueller_20_mhd,Varma2023, Sykes2024}.

Alongside these improvements, it became clear already in the 1990s that 1D simulations of CCSNe were insufficient to capture 
key aspects of the explosion mechanism, and multidimensional models were needed \citep{Herant1995,Burrows1995,janka_96}. We now know that 
even 2D simulations are insufficient, and we need full 3D simulations over long timescales \citep{Janka2025, mueller_15_3d,Burrows2024}
to capture the complexity of the CCSN mechanism. 

The need for multi-dimensional simulations, however, is not just limited to the explosion mechanism.
Over a decade ago, \citet{Couch2013} and \citet{Muller2015b} showed that symmetry breaking by 
the convective flow in the final phases of stellar evolution is crucial for the explosion mechanism.

Convective motions, primarily in the oxygen and silicon shells of massive stars, generate large-scale velocity and entropy perturbations that are advected during collapse. As these perturbations fall toward the core, the acceleration and compression of the infalling material partially convert vorticity and entropy fluctuations into acoustic waves through vortical–acoustic coupling \citep{Foglizzo2001, Abdikamalov2016, Telman2024}. When these perturbations reach the stalled accretion shock, the combined acoustic and vortical forcing excites shock oscillations and strong post-shock turbulence. This results in additional turbulent pressure in the gain region that acts on the stalled shock \citep{Abdikamalov2018}.

\citet{Muller2015b} and \citet{Mueller_2016} showed that these perturbations aid the explosion mechanism by reducing the critical luminosity required for a successful shock revival and CCSN explosion:

\begin{align}
\frac{\Delta L_\mathrm{crit}}{L_\mathrm{crit}} \sim \frac{4.7 \mathcal{M}_\mathrm{conv}}{\ell_\mathrm{conv}},
\label{eq:Lcrit}
\end{align}
Where $L_\mathrm{crit}$ is the critical luminosity required to successfully drive a CCSN explosion, $\mathcal{M}_\mathrm{conv}$ is the Mach number of the convective flow and $\ell_\mathrm{conv}$ is the angular wavenumber scale of the convective plumes. In particular, large magnitudes ($\mathcal{M} \gtrsim 0.1$) and large scales ($\ell \lesssim  4$) of the pre-collapse convective flows are some of the most important features that determine whether these seed perturbations can help facilitate shock revival. Analytic and numerical studies show that this process lowers the critical neutrino luminosity for explosion by roughly 10–25\% depending on the convective Mach number and perturbation scale \citep{Muller2015b, Mueller_2017, Kazeroni2020, Abdikamalov2020}.

While a number of simulations have followed the final minutes of convection in 3D \citep{Couch_2015, Mueller_2016, Yadav2020, Yoshida2021},
fewer still have followed the ensuing explosion \citep{Mueller_2017, bollig_21} using these 3D progenitor structure.
Due to the high computational costs of these simulations, the vast majority of CCSN simulations still use 
1D stellar evolution models as initial conditions to their 3D CCSN simulations, breaking the spherical symmetry 
by either using simplified perturbations to the initial velocities and/or density profiles \citep[e.g.,][]{Buras2006, Just2018}.

Several methods have been proposed to date to generate multi-dimensional structure that resembles 
convection, such as using scalar spherical harmonics in 2D \citep{Muller2015b}. Both \citep{Chen2013} 
and \citet{Chatzopoulos2015} suggested methods to reproduce the turbulent spectrum of convection in 3D.
However, the method of \citet{Chen2013} imposed a turbulent velocity field by performing a Fourier decomposition 
of a Kolmogorov turbulence spectrum, which is no longer applicable at the largest scales, and did not take constraints on the velocity field for subsonic flow into account. Instead of constructing a purely artificial field, \citet{Chatzopoulos2015} 
proposed a method to analyse stochastic 3D stellar velocity fields by decomposing the 
velocity field from multi-dimensional simulations into vector spherical harmonics (VSH). Although this method
allows a reconstruction of a realistic velocity field when initialising CCSN simulations, 
it still depends on data from expensive 3D stellar convection simulations, and the generalisation of the spectral properties to new progenitor models is unclear.

In this work, we present a method to generate synthetic 3D velocity fields of the dominant modes in convective zones from 1D stellar evolution profile inputs. The method uses the location and ratios of the convective shell boundaries to approximately determine the dominant convective scale, which is also typically similar in scale to the most unstable mode in the linear regime \citep{Chandrasekhar1961, Foglizzo2006}, even in the regime of fully developed non-linear convection \citep{Mueller_2016}. Using VSH and physical arguments pertaining to stratified, buoyancy-driven convection, we present a mathematical formulation to construct a 3D velocity field. This formulation is presented as an open \textsc{Python} code that reads data from \texttt{MESA} \citep{Paxton2011, Paxton2019}, and generates the 3D velocity fields of regions that are expected to be convectively unstable. The velocity field is also automatically scaled to velocities predicted by mixing-length theory (MLT) in these unstable regions.

While this method does not yet reproduce a full turbulent spectrum, it reproduces both the magnitude and largest scale of the perturbations, which are the key properties necessary for CCSN simulations. This can then be used as 3D initial conditions for CCSN simulations, forgoing the need for expensive 3D stellar convection simulations, while still capturing the essential features of the convective flow. 

In Section~\ref{sec:theory}, we describe the theoretical framework of our method, and in Section~\ref{sec:implementation}, 
we describe the practical constraints and choices we make in implementing our method. Finally, we present 
some examples of the generated velocity fields, both idealised and generated from 1D \texttt{MESA} models 
in Section~\ref{sec:velocity_field} before concluding (Section~\ref{sec:conclusion}). 

\section{Theory}
\label{sec:theory}
To construct synthetic velocity fields that can act as realistic initial conditions for core-collapse supernova simulations, we need a mathematical framework that both reflects the geometry of convective flows and enforces key physical constraints. A natural basis for such a description is provided by vector spherical harmonics (VSH), which allow us to decompose velocity perturbations into well-defined angular modes. By combining the VSH expansion with the Helmholtz decomposition, along with physical constraints (e.g., anelastic flow, non-radial vorticity), we can simplify our system of equations and arrive at a compact prescription for the velocity field. We introduce the relevant definitions and develop the formalism that underpins our method as follows.

We begin by defining the VSH, $\mathbf{Y}_{\ell}^{m}$, $\mathbf{\Psi}_{\ell}^{m}$ 
and $\mathbf{\Phi}_{\ell}^{m}$  using the scalar spherical harmonics $Y_{\ell}^{m}$ of degree (angular wavenumber) $\ell$ 
and order $m$ \citep{Arfken2005, Jackson1998}:

\begin{align}
    \mathbf{Y}_{\ell}^{m}&=
     Y_{\ell}^{m} \mathbf{e}_r,
     \\
     \mathbf{\Psi}_{\ell}^{m}&=
     r \nabla Y_{\ell}^{m}\\
     &=
     \left(m \cot \theta  Y_\ell^m +
     e^{-i \varphi } \sqrt{(1+\ell-m)(\ell+m)} Y_\ell^{m+1}
     \right)
     \mathbf{e}_\theta+
     \frac{i m r}{\sin \theta}  Y_{\ell}^{m} \mathbf{e}_\varphi,
     \nonumber\\
     \mathbf{\Phi}_{\ell}^{m}&=
     \mathbf{r}\times \nabla Y_{\ell}^{m}\\
     &=
     \left(m \cot \theta  Y_\ell^m +
     e^{-i \varphi } \sqrt{(1+\ell-m)(\ell+m)} Y_\ell^{m+1}
     \right)\mathbf{e}_\varphi
     -\frac{i m r}{\sin \theta}  Y_{\ell}^{m} \mathbf{e}_\theta,
     \nonumber
\end{align}
where $\mathbf{e}_r$, $\mathbf{e}_\theta$ and $\mathbf{e}_\varphi$ are the unit vectors in each direction in spherical polar coordinates.

For simplicity, we assume that the fluid in all stellar convective zones we are considering adheres to the anelastic approximation, such that $\nabla \cdot \rho \mathbf{v} = 0$,
and the deviations of the density field from the spherical background stratification remain small. This is well motivated since, even in the highest velocity cases, the interior convective zones in massive stars usually have Mach numbers ($\mathcal{M}$) of the order $\mathcal{M}\, \approx 0.1$. Using the relation $\delta \rho\approx \rho v^2/c_\mathrm{s}^2$ (see section 10 of \citet{Landau1959}), we see that density fluctuations in these regions are therefore $\delta \rho/\rho \approx \mathcal{M}^{2} \approx 10^{-2}$.

We apply the Helmholtz decomposition (e.g. \citet{Jackson1998}) to our velocity field, separating it into its solenoidal and irrotational components,

\begin{align}
    \rho \mathbf{v} = \rho (\nabla \times \mathbf{A} + \nabla B). 
\end{align}
Our assumption of anelastic flow allows us to keep just the solenoidal (curl) part of the equation:
\begin{align}
\rho \mathbf{v} = \nabla \times \rho\mathbf{A}.
\end{align}
We find the curl of each of the three families of VSH, and weigh them by the set of radial basis functions $f(r)$, $g(r)$ and $h(r)$, 

\begin{align}
    \nabla \times (f(r) \mathbf{Y}_{\ell}^{m})&=\frac{f(r)}{r}\mathbf{\Phi}_\ell^m,
    \label{eq:curl_y}
     \\
     \nabla \times (g(r)\mathbf{\Psi}_{\ell}^{m})&=\left(g'(r) +\frac{g (r)}{r}\right) \mathbf{\Phi}_\ell^m,
     \label{eq:curl_psi}
     \\
     \nabla \times (h(r) \mathbf{\Phi}_{\ell}^{m})&=-\frac{h(r) \ell (\ell+1)}{r} \mathbf{Y}_{\ell}^{m}
    -\left(h'(r)+\frac{h(r)}{r}\right) \mathbf{\Psi}_{\ell}^{m},
    \label{eq:curl_phi}
\end{align}
and use these to define the momentum vector field,

\begin{align}
    \rho \mathbf {v}
    &=
    \nabla \times (f(r) \mathbf{Y}_{\ell}^{m}+g(r) \mathbf{\Psi}_{\ell}^{m}+h(r) \mathbf{\Phi}_{\ell}^{m})
    \nonumber
    \\
    &=\frac{f(r)}{r }\mathbf{\Phi}_\ell^m+
    \left(g'(r) +\frac{g (r)}{r}\right) \mathbf{\Phi}_\ell^m
    -\frac{h(r) \ell (\ell+1)}{r} \mathbf{Y}_{\ell}^{m}
    \nonumber
    \\
    &\phantom{=}
    -\left(h'(r)+\frac{h(r)}{r}\right) \mathbf{\Psi}_{\ell}^{m}.
    \label{eq:flow_field1}
\end{align}
Equation~(\ref{eq:flow_field1}) is the most general solenoidal momentum field we can construct with VSH, which we use as our starting point.

\subsection{The Requirement of Non-Radial Vorticity}

In subsonic, stratified, buoyancy-driven convection (without rotation), vorticity ($\boldsymbol{\omega}=\nabla \times \mathbf{v}$) is primarily generated by baroclinicity. This can be best seen in the vorticity transport equation (Equation~\ref{eq:vorticity_t}): 

\begin{equation}
    \frac{\ud \boldsymbol{\omega}}{\ud t}
    =
    \underbrace{(\boldsymbol{\omega}\cdot\nabla) \mathbf{v}}_{a}
    -\underbrace{\boldsymbol{\omega} (\nabla \cdot\mathbf{v})}_{b}
    +\underbrace{\frac{\nabla\rho \times \nabla P}{\rho^2}}_{c},
    \label{eq:vorticity_t}
\end{equation}
where buoyant driving enters via the baroclinic term, $c$. For a stratified background, as is the case for stellar interiors, the pressure effectively depends on gravity alone, and we assume that pressure perturbations, $P'$, due to other accelerations are negligible. This means that only the radial component of $\nabla P$ survives, and the baroclinic term is purely \emph{non-radial}. For an incompressible field, the term $b$ disappears, and the term generally does not change the direction of the vorticity generated by the baroclinic term. The term $a$ is the vortex stretching term, which does not generate large coherent $\omega$ components \citep{Turner1980, Chassaing2002}. 

This places a strong constraint on our velocity field, as we now require the vorticity of our constructed field to also be purely non-radial. Calculating the vorticity of Equation~(\ref{eq:flow_field1}), we find

\begin{align}
    \boldsymbol{\omega}&=\nabla \times \mathbf{v}
    =
    \nabla \times \left[\left(\frac{f(r)}{r \rho(r) }+
    \frac{g'(r)}{\rho(r)} +\frac{g (r)}{r \rho(r)}\right) \mathbf{\Phi}_\ell^m\right]
    \label{eq:vorticity}
\\ 
&\phantom{=}
    -\nabla \times \left(\frac{h(r) \ell (\ell+1)}{r \rho(r)} \mathbf{Y}_{\ell}^{m}\right)
    -\nabla \times \left(
    \left(\frac{h'(r)}{\rho(r)}+\frac{h(r)}{r \rho(r)}\right) \mathbf{\Psi}_{\ell}^{m}\right).
    \nonumber
\end{align}

We break down Equation~(\ref{eq:vorticity}) further by introducing $X(r)=[(f(r)+r g'(r)+g(r))/(r\rho(r))]$
and using Equation~(\ref{eq:curl_phi}), to find that the first term of Equation~(\ref{eq:vorticity}) involving the functions $f$ and $g$, gives rise to both a radial and a non-radial component of the vorticity,

\begin{align}
     \nabla \times \left(X(r)\mathbf{\Phi}_\ell^m\right)
     =
     &-\frac{X(r) \ell (\ell+1)}{r} \mathbf{Y}_{\ell}^{m}
    -\left(X'(r)+\frac{X(r)}{r}\right) \mathbf{\Psi}_{\ell}^{m}.
    \label{eq:vorticity_families_1_and_2}
\end{align}

On the other hand, by virtue of Equations~(\ref{eq:curl_y})
and (\ref{eq:curl_psi}), the terms containing $h$ only give rise
to non-radial vorticity components proportional to $\mathbf{\Phi}_\ell^m$.

Therefore, if the radial vorticity is to vanish everywhere, this requires
$X(r)\equiv 0$ or $f(r)\equiv r g'(r)+g(r)$. However, in this case,
the term proportional to $\mathbf{\Psi}_\ell^m$ in Equation~(\ref{eq:vorticity_families_1_and_2}) also vanishes, and in fact,
all the terms containing $f$ and $g$ in the momentum
density cancel. In other words, one can simply construct single-mode stream functions from  $\mathbf{\Phi}_\ell^m$ alone. This remains true for stream functions for multiple modes with different $\ell$ and $m$ because the radial vorticity component will only vanish identically everywhere if it does so for any terms in the orthogonal expansion in different spherical harmonics.

More simply, using the relations for the curl of the vector spherical harmonics, one immediately finds that the terms containing $h$ in Equation~(\ref{eq:vorticity}) only give rise to non-radial vorticity terms proportional to $\mathbf{\Phi}_\ell^m$. The requirement of a non-radial vorticity field implies that only one family of vector spherical harmonics needs to be taken into account in the stream function. This reduces our momentum field to:

\begin{tcolorbox}[standard jigsaw, opacityback=0]
\begin{align}
    \rho \mathbf {v}
    &= 
     \nabla \times \left(h(r)\mathbf{\Phi}_\ell^m\right)
    \label{eq:flow_field}
    \\
    &=
    -\frac{h(r) \ell (\ell+1)}{r} \mathbf{Y}_{\ell}^{m}
    \nonumber  -\left(h'(r)+\frac{h(r)}{r}\right) \mathbf{\Psi}_{\ell}^{m}.
\end{align}
\end{tcolorbox}

Given a reasonable choice for the basis function $h(r)$, Equation~(\ref{eq:flow_field}) now gives us a physically motivated velocity/momentum field suitable to represent buoyancy-driven flows in stellar interiors. 

\subsection{Angular Momentum}
A final physical consideration for Equation~(\ref{eq:flow_field}) before we can apply it is that the synthetic convective flow should not unintentionally introduce bulk rotational flow. It is easy to verify that this is not a problem with any modes constructed from $\mathbf{\Phi}_{\ell}^{m}$ as the total angular momentum contained in these modes vanishes. 

To see this, consider the angular momentum density,
\begin{align}
    \mathbf{r}\times \rho\mathbf{v}=
    -\left(h'(r)+\frac{h(r)}{r}\right) \mathbf{r}\times \mathbf{\Psi}_{\ell}^{m}=
    -\left(h'(r)+\frac{h(r)}{r}\right) \mathbf{\Phi}_{\ell}^{m}.
    \label{eq:ang_mom}
\end{align}

The total angular momentum can be obtained by dotting
$\mathbf{r}\times \rho\mathbf{v}$ with the unit vectors
along the $x$-, $y$-, and $z$- axis to extract the corresponding
components of the angular momentum. Expressing these
units vectors in vector spherical harmonics
\begin{align}
\mathbf{e}_x&=\sqrt{2\pi/3} (\mathbf{Y}_1^{-1}-\mathbf{Y}_1^1
+\mathbf{\Psi}_1^{-1}-\mathbf{\Psi}_1^1),
\\
\mathbf{e}_y&=i\sqrt{2\pi/3} (\mathbf{Y}_1^{-1}+\mathbf{Y}_1^1
+\mathbf{\Psi}_1^{-1}+\mathbf{\Psi}_1^1),
\\
\mathbf{e}_z&=2\sqrt{\pi/3} (\mathbf{Y}_1^0+\mathbf{\Psi}_1^0),
\end{align}
one finds that the total angular momentum components are
just projections of the mode onto vector spherical harmonics 
$\mathbf{Y}_1^m$ and $\mathbf{\Psi}_1^m$
of degree $\ell=1$, e.g.,
\begin{align}
L_z&=\int e_\mathbf{z }\cdot (\mathbf{r}\times \rho\mathbf{v}) \,\ud V
\\
&=
-\int 2 \sqrt{\frac{\pi }{3}} (
\mathbf{Y}_1^0+\mathbf{\Psi}_1^0)
\left(h'(r)+\frac{h(r)}{r}\right) \mathbf{\Phi}_{\ell}^{m}\,\ud V=0.
\end{align}
This integral, like those for $L_x$ and $L_y$, vanishes because of the orthogonality $\mathbf{\Phi}_{\ell}^{m}$  with any of the vector spherical harmonics of the other two families.

\begin{figure}
    \centering
    \includegraphics[width=\linewidth]{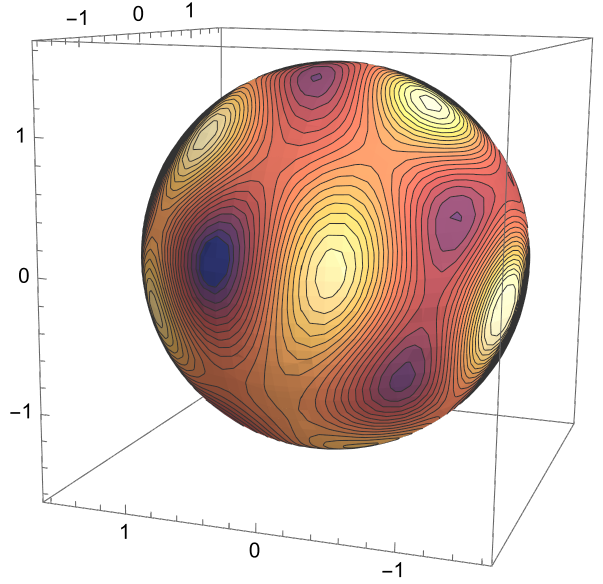}
    \caption{Contour plot of radial velocity (in arbitrary units) for a random superposition of modes with $\ell=6$
    on a sphere of radius $1.5$ in the middle of a convective shell between $R_\mathrm{i}=1$ and $R_\mathrm{o}=2$.}
    \label{fig:random_vr}
\end{figure}

\section{Implementation}
\label{sec:implementation}
From our discussion in Section~\ref{sec:theory}, Equation~(\ref{eq:flow_field}) gives us an equation for the momentum field that satisfies our imposed physical conditions (i.e. anelastic flow with non-radial vorticity and zero net angular momentum). 
However, we are still left with some choices to make when initialising the flow field to ensure that the velocity field resembles realistic convective flow patterns.

\subsection{Multi-mode Flow}
Equation~(\ref{eq:flow_field}) shows that for single modes, the radial velocity pattern of convective updrafts and downdrafts  is given directly by the underlying scalar spherical harmonic $Y_\ell^m$. This is problematic because it makes the flow highly anisotropic. For example, for $m=0$, the flow is axisymmetric, and for $l=\pm m$, the flow is mostly confined to low latitudes near the equator. For some values of $m$, the shape of single-mode convective cells will also be far from isotropic and more sheet-like, rather than the exhibiting the relatively roundish shape of convective plumes in realistic convection. Contrary to the 2D case \citep{Muller2015b}, single-mode flow patterns are therefore inappropriate in 3D.

In order to ensure statistical isotropy and roundish shapes of the updrafts and downdrafts while retaining a definite angular wavenumber $\ell$, we populate modes of different $m$ with randomly distributed amplitudes and $A_m$ and phases $\phi_m$. The stream function is expressed as
\begin{equation}
    \mathbf{\psi}
    =h(r) \,\mathrm{Re} \sum_{m=-\ell}^{\ell} A_{m} e^{i\phi_m} \mathbf{\Phi}_{\ell}^{m},
    \label{eq:psi}
\end{equation}
where $A_m$ is normally distributed with unit variance and $\phi_m$ follows a uniform distribution in the interval $[0,2\pi]$. The resulting radial velocity field is illustrated for one particular realisation with $\ell=6$ in 
Figure~\ref{fig:random_vr}. The figure shows that the updrafts and downdrafts have roughly equal areas, roundish shapes, and are isotropically distributed across the sphere.

In the future, more modes with different spherical harmonics degree $\ell$, and also modes with different radial dependence could be included to better mimic the realistic turbulence spectrum of buoyancy-driven convection. In addition, the assumption of a Gaussian amplitude distribution should be investigated and, if need be, based on 3D simulation data.

\begin{figure}
    \centering
    \includegraphics[width=\linewidth]{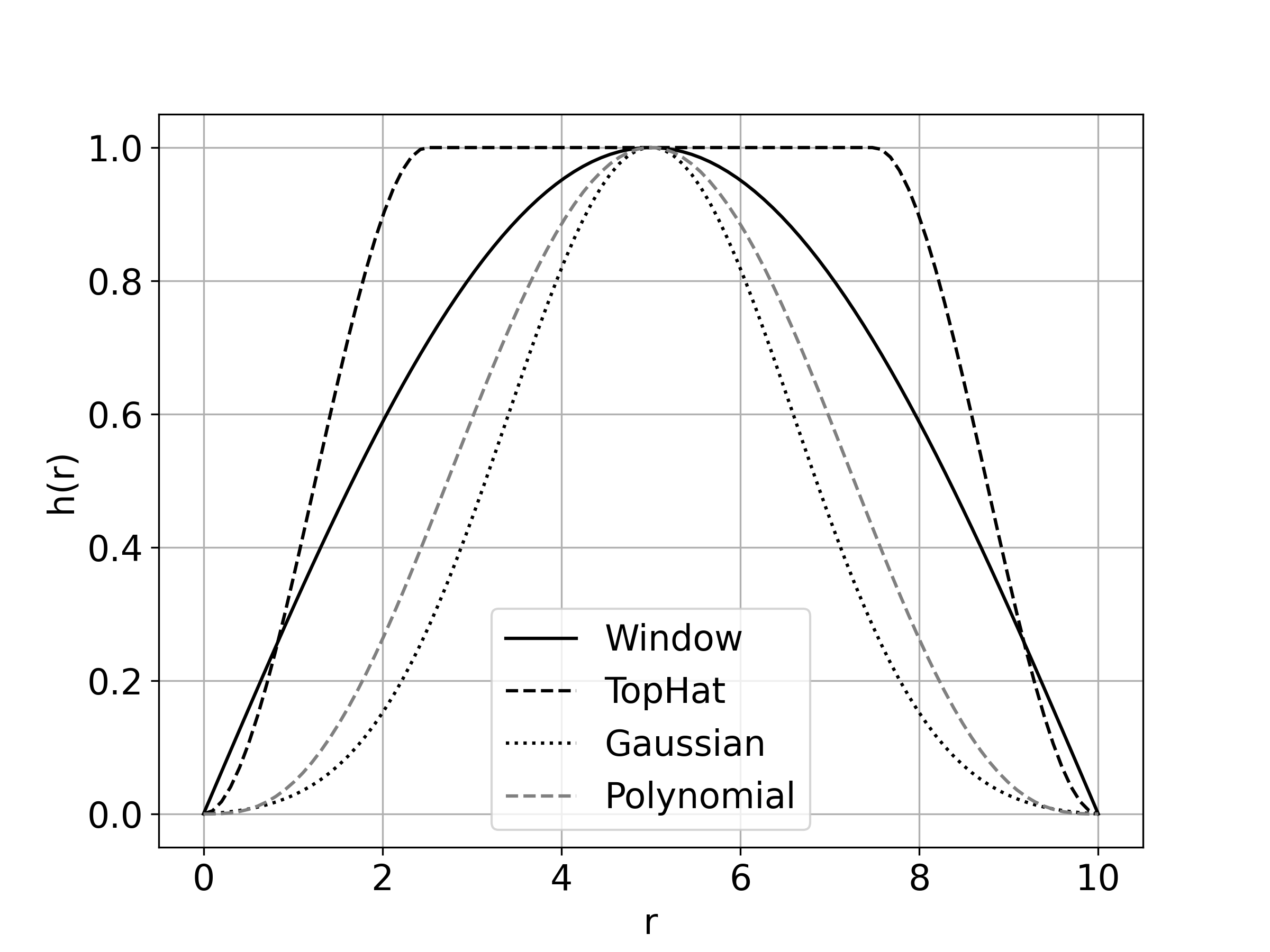}
    \caption{Plot visualising the different options for radial basis functions currently implemented.}
    \label{fig:basis_func}
\end{figure}

\subsection{Radial Basis Functions}
After imposing the constraint of non-radial vorticity on our model in Section~\ref{sec:theory}, our equations 
simplify considerably, and we lose the dependence on the radial basis functions $f(r)$ and $g(r)$.
The remaining radial basis function $h(r)$ is still an unconstrained function that can be chosen to represent the radial dependence of the convective flow in stars. They must, however, be subject to appropriate boundary conditions: To ensure that the radial velocity drops to zero at the inner and outer boundary, $h(r)$ must vanish there (or drop very quickly outside the boundaries if we allow for some amount of overshooting).

We find that the precise choice of $h(r)$ is not critical qualitatively, and likely depends on the specific case being explored. 
The important constraint on the radial basis function is that the function must be positive and non-zero in the region of interest, 
and vanishes at the radial boundaries.

We show four different radial basis functions that fit these constraints and have been implemented in our code, shown in Figure~\ref{fig:basis_func}, shown in arbitrary units. The functions we show are as follows: 

\noindent 
1) Windowed sine
\begin{align}
    h_{\mathrm{window}}(r)=
    \sin\,\!\Bigl(\pi\,\frac{r-R_\mathrm{i}}{R_\mathrm{o}-R_\mathrm{i}}\Bigr), 
    \label{eq:window}
\end{align}
2) Gaussian
\begin{align}
    h_{\mathrm{gauss}}(r)=
    4\,\exp\,\!\Bigl(-\bigl(\tfrac{r-r_0}{\sigma}\bigr)^{\!2}\Bigr)\,f(r),
    \label{eq:gauss}
\end{align}
where $r_0$, $\sigma$ and the function, $f(r)$ are defined as follows,
\begin{align}%
    r_0=\frac{R_\mathrm{i}+R_\mathrm{o}}{2},\qquad
    \sigma=\frac{R_\mathrm{o}-R_\mathrm{i}}{4},\qquad
    f(r)=\frac{(r-R_\mathrm{i})(R_\mathrm{o}-r)}{(R_\mathrm{o}-R_\mathrm{i})^2}, \nonumber
\end{align}
3) Polynomial (with $n=3$)
\begin{align}
    h_{\mathrm{poly}}(r)=
    \Biggl(\frac{4\,(r-R_\mathrm{i})(R_\mathrm{o}-r)}{(R_\mathrm{o}-R_\mathrm{i})^2}\Biggr)^{n},
    \label{eq:poly}
\end{align}
4) Top Hat
\begin{align}
    S(t)=3t^2-2t^3,\qquad \delta=\frac{R_\mathrm{i}+R_\mathrm{o}}{4}, \nonumber
    \\
    h_{\mathrm{top}}(r)=
    \begin{cases}
    0, & r<R_\mathrm{i},\\[4pt]
    S\,\!\Bigl(\dfrac{r-R_\mathrm{i}}{\delta}\Bigr), & R_\mathrm{i}\le r< R_\mathrm{i}+\delta,\\[10pt]
    1, & R_\mathrm{i}+\delta \le r \le R_\mathrm{o}-\delta,\\[6pt]
    S\,\!\Bigl(\dfrac{R_\mathrm{o}-r}{\delta}\Bigr), & R_\mathrm{o}-\delta < r \le R_\mathrm{o},\\[10pt]
    0, & r>R_\mathrm{o}~,
    \label{eq:top_hat}
    \end{cases}
\end{align}
where the function $S(t)$ is a smoothing function to prevent sharp changes. All of these functions follow the conditions that, $h(r)\ge 0$, $R_\mathrm{i}\le r\le R_\mathrm{o}$ and $h(R_\mathrm{i})=h(R_\mathrm{o})=0$, where $R_\mathrm{i}$ and $R_\mathrm{o}$ are the radii of the inner and outer boundaries.

Some of these functions, namely the windowed sine function and top-hat function, are chosen as they resemble profiles that have been observed in 3D simulations \citep{Cristini2017, Georgy2024}.
Specifically, the windowed sine function resembles the radial velocity profiles, and the top-hat function resembles the profile of the total velocity field. The MLT prescription in 1D stellar evolution also often produces predictions that are consistent with a top-hat profile. We note that due to the large density gradients within a convective shell, we have also implemented equivalent density weighted functions for Equations~\ref{eq:window} -~\ref{eq:top_hat}.

These additional constraints on our field now allow us to construct plausible 3D convective velocity fields.

\begin{figure*}
    \centering
        \begin{subfigure}{0.45\linewidth}
            \includegraphics[width=\linewidth]{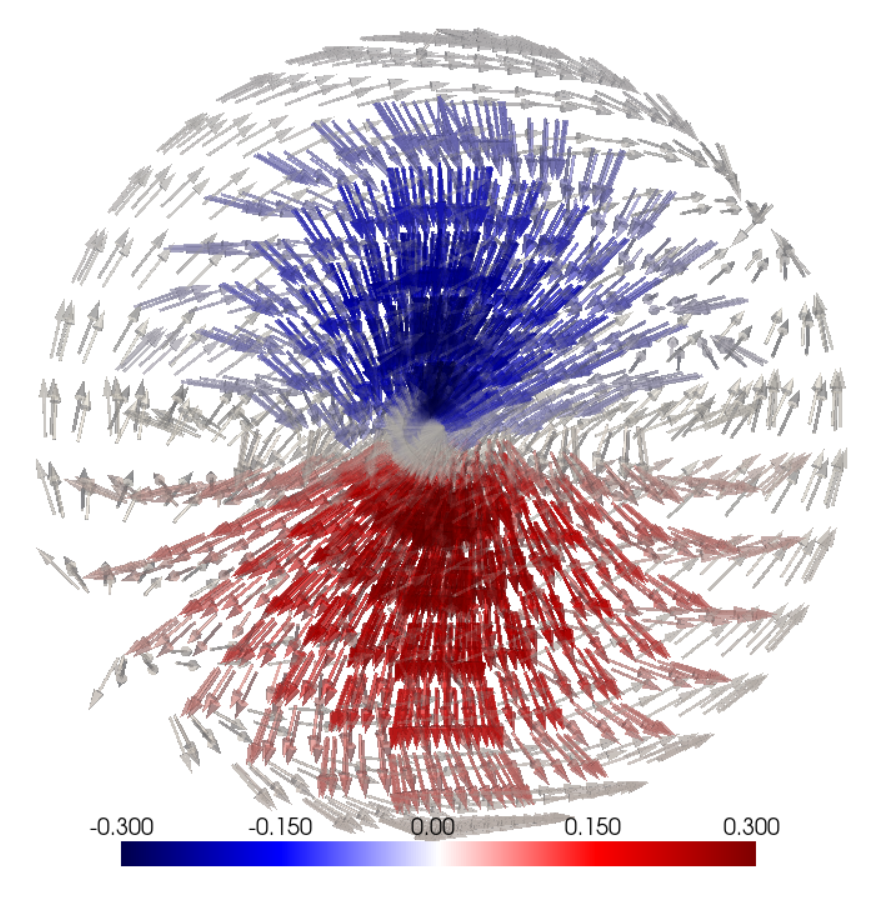}
        \end{subfigure}
    \hfil
        \begin{subfigure}{0.45\linewidth}
            \includegraphics[width=\linewidth]{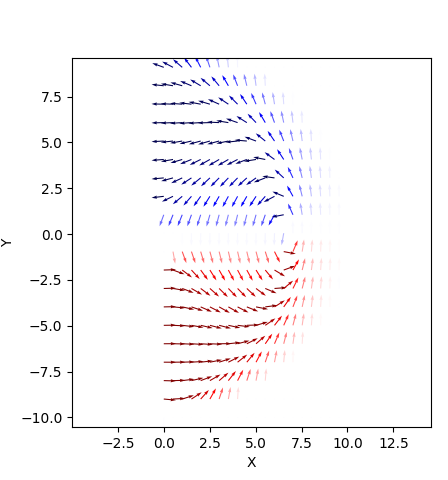}
        \end{subfigure}
    
    \caption{3D vector plot (left) and 2D vector slice for a dimensionless dipole velocity field configuration ($\ell$ = 1). The vector arrows include all three velocity components, but the colourbar indicates just the radial velocity.}
        \label{fig:vector_dipole}
\end{figure*}
    
\section{Velocity Field Generation}
\label{sec:velocity_field}

The velocity field prescription described in the previous sections has been implemented in \textsc{Python}, and is available for use on GitLab \footnote{\url{https://gitlab.com/vishnuvarma/Stellar_velocity_field}}.
Figure~\ref{fig:vector_dipole} shows an example of the 3D vector plot and a 2D slice of the velocity field for a dipole configuration ($\ell$ = 1) in arbitrary units.
The 3D vector plot is coloured by the radial velocity component, with vector arrows showing the full velocity field. 
The code uses an approximation of the angular wavenumber $\ell$ to generate the velocity field using the ratio of the inner and outer radius of the convective shell, 
$R_\mathrm{i}$ and $R_\mathrm{o}$:
\begin{align}
    \ell = \frac{\pi (R_\mathrm{o} + R_\mathrm{i})}{2(R_\mathrm{o} - R_\mathrm{i})}.
    \label{eq:ell_approx}
\end{align}
This is a simple estimate that corresponds to the fastest growing mode in linear stability analysis,
and has been shown to be largely compatible with exact calculations \citep{Chandrasekhar1961,Foglizzo2006}.

As we will show, a simple dimensionless velocity field can be generated using $R_\mathrm{i}$ and $R_\mathrm{o}$ in arbitrary units. However, for a more realistic use case, $R_\mathrm{i}$ and $R_\mathrm{o}$, and hence, the angular wavenumber $\ell$ can be determined directly from output data from stellar evolution models. We have implemented this for the open-source \texttt{MESA} code, but it is easy to modify the code to obtain velocity fields from other stellar evolution codes.

\subsection{Dimensionless Velocity Field}
\begin{figure*}
\centering
    \begin{subfigure}{0.33\linewidth}
        \includegraphics[width=\linewidth]{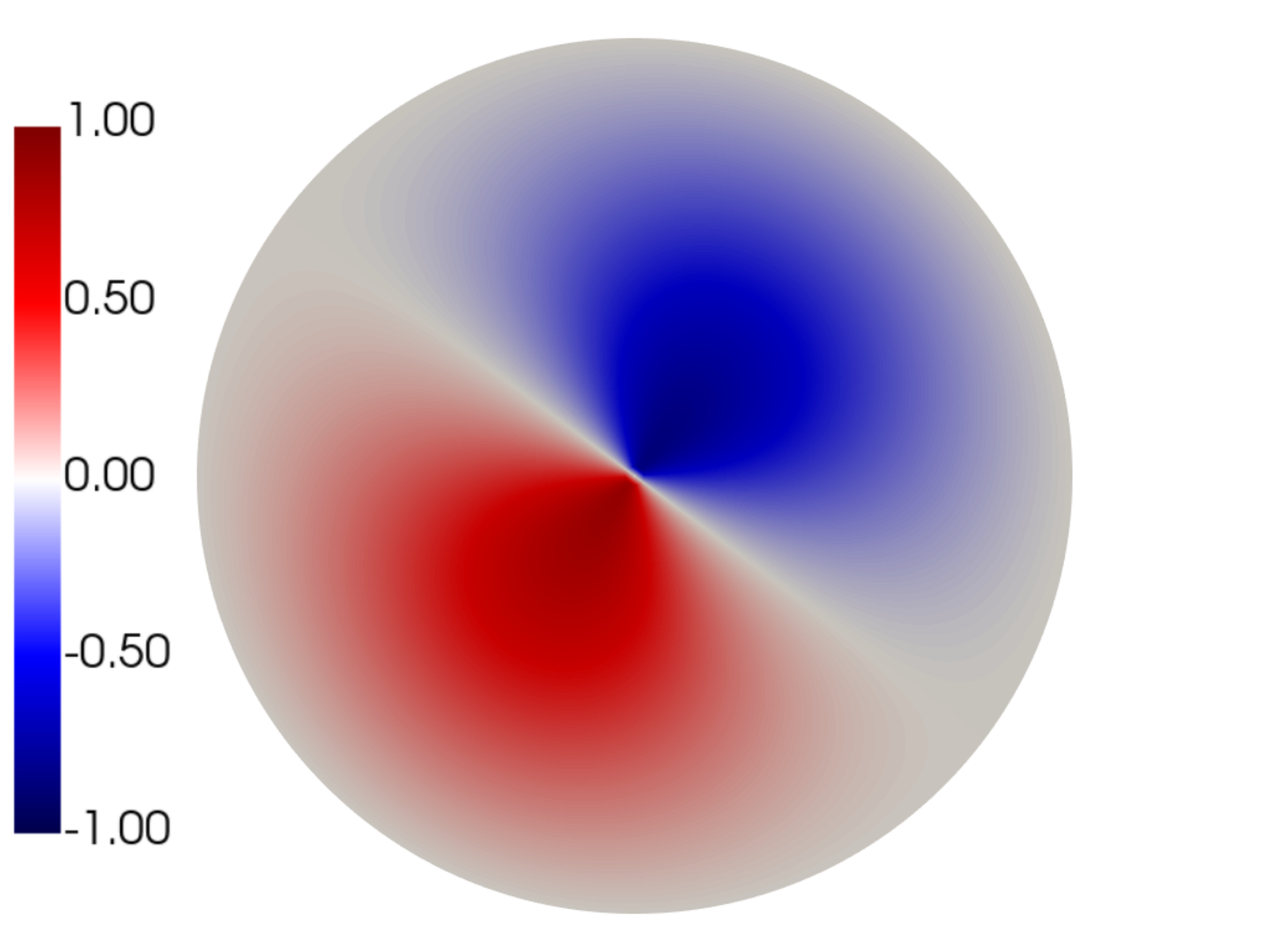}
    \end{subfigure}
\hfil
    \begin{subfigure}{0.33\linewidth}
        \includegraphics[width=\linewidth]{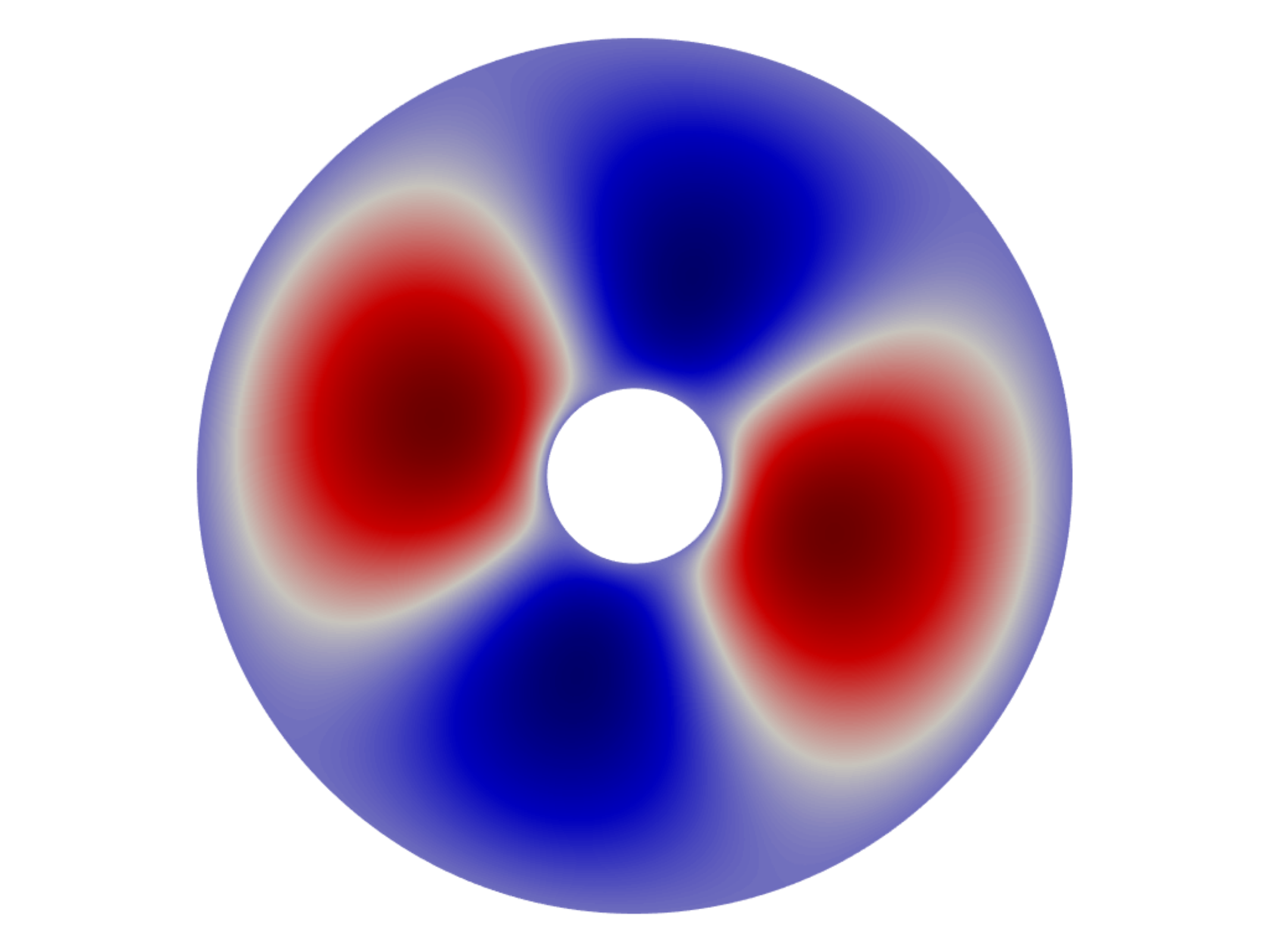}
    \end{subfigure}
\hfil
    \begin{subfigure}{0.33\linewidth}
        \includegraphics[width=\linewidth]{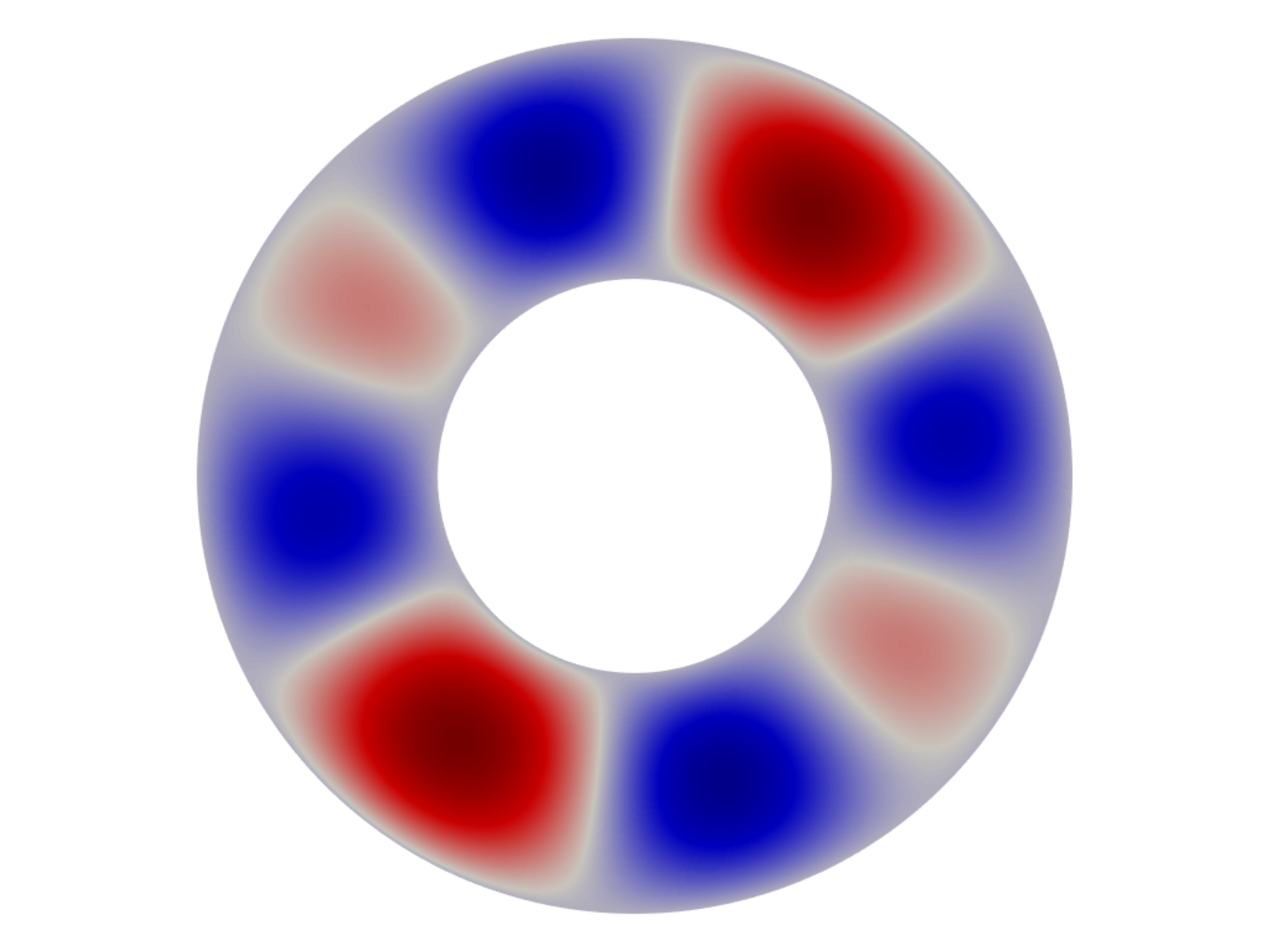}
    \end{subfigure}

    \begin{subfigure}{0.33\linewidth}
        \includegraphics[width=\linewidth]{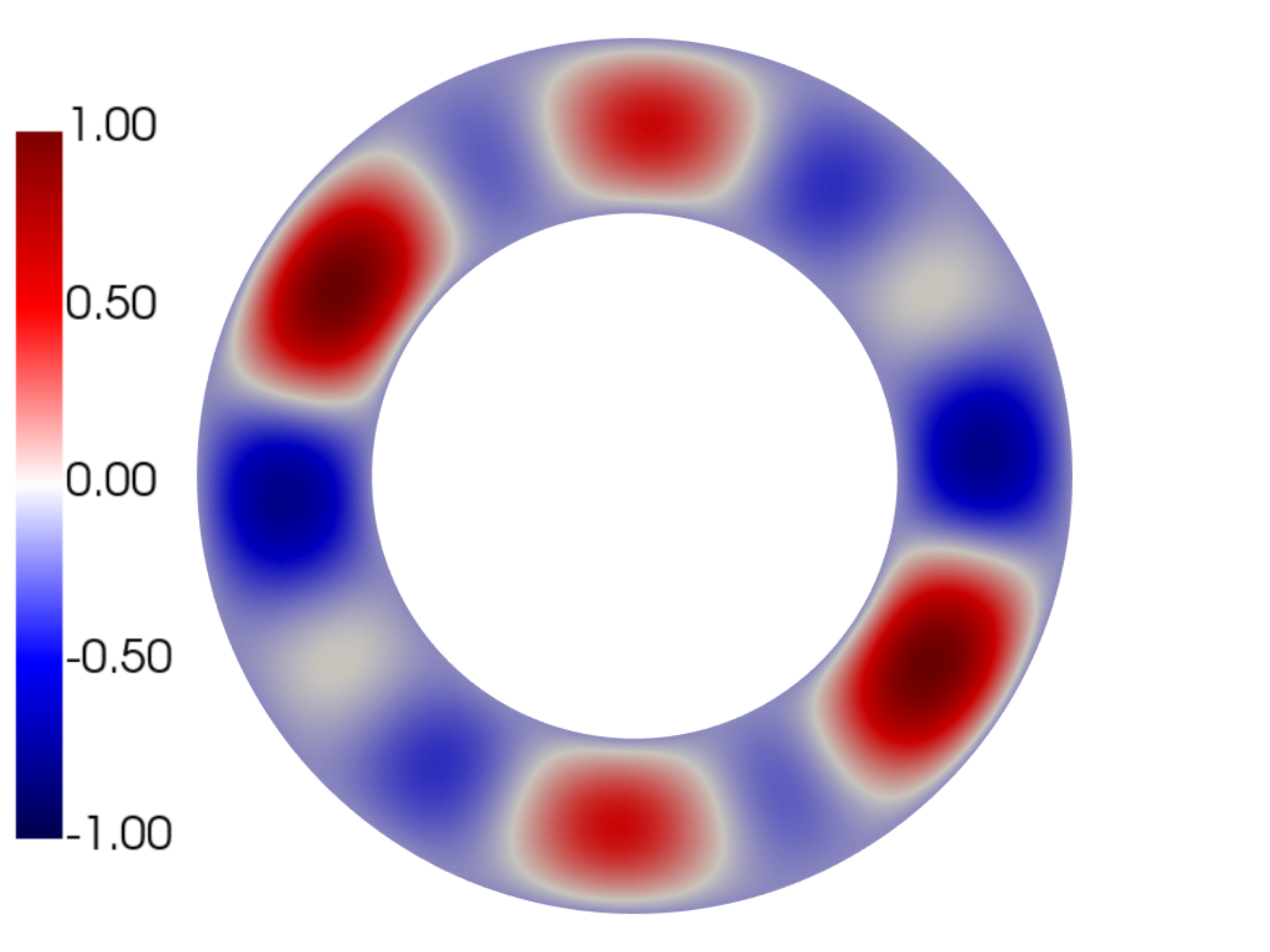}
    \end{subfigure}
\hfil
    \begin{subfigure}{0.33\linewidth}
        \includegraphics[width=\linewidth]{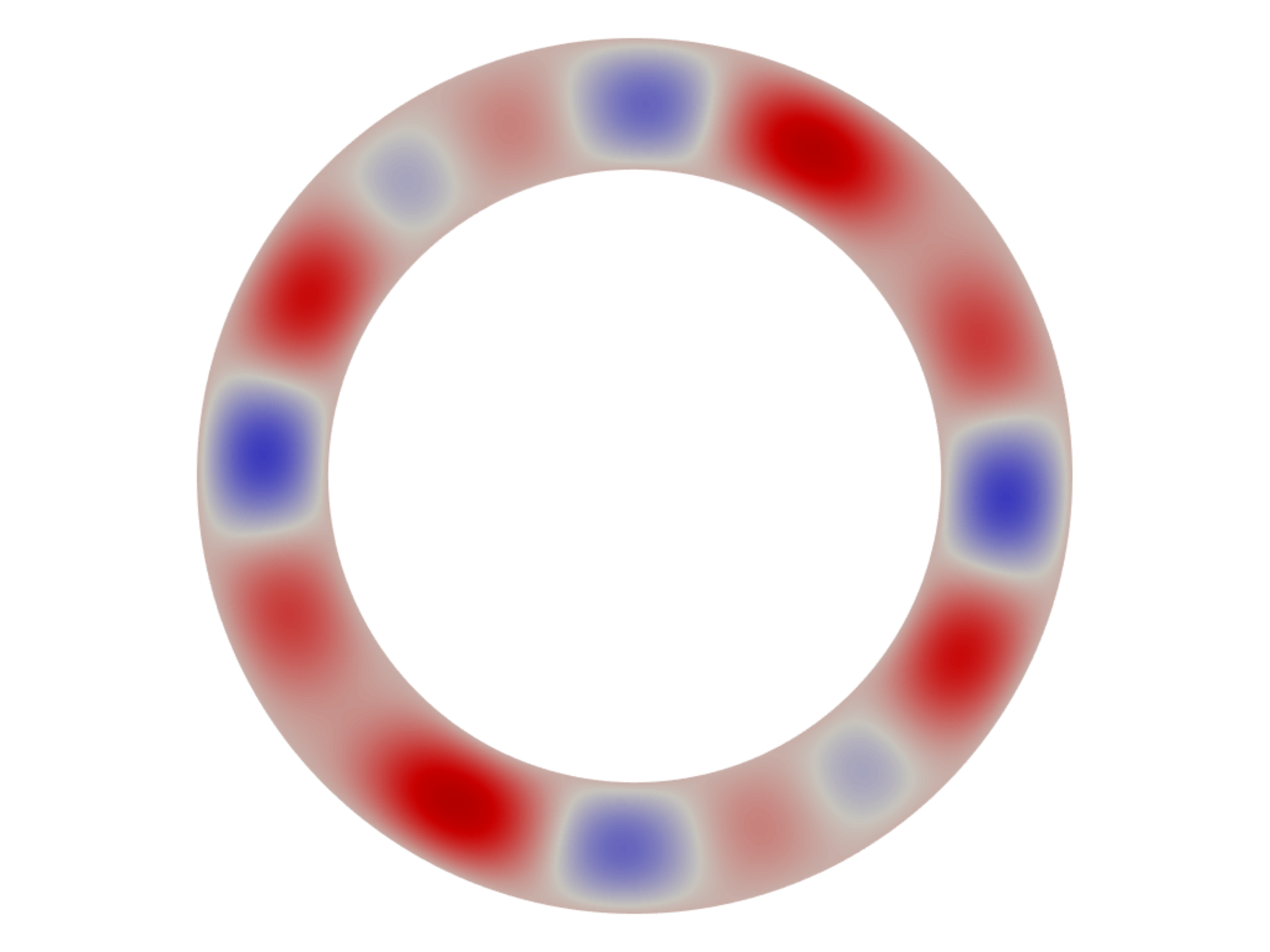}
    \end{subfigure}
\hfil
    \begin{subfigure}{0.33\linewidth}
        \includegraphics[width=\linewidth]{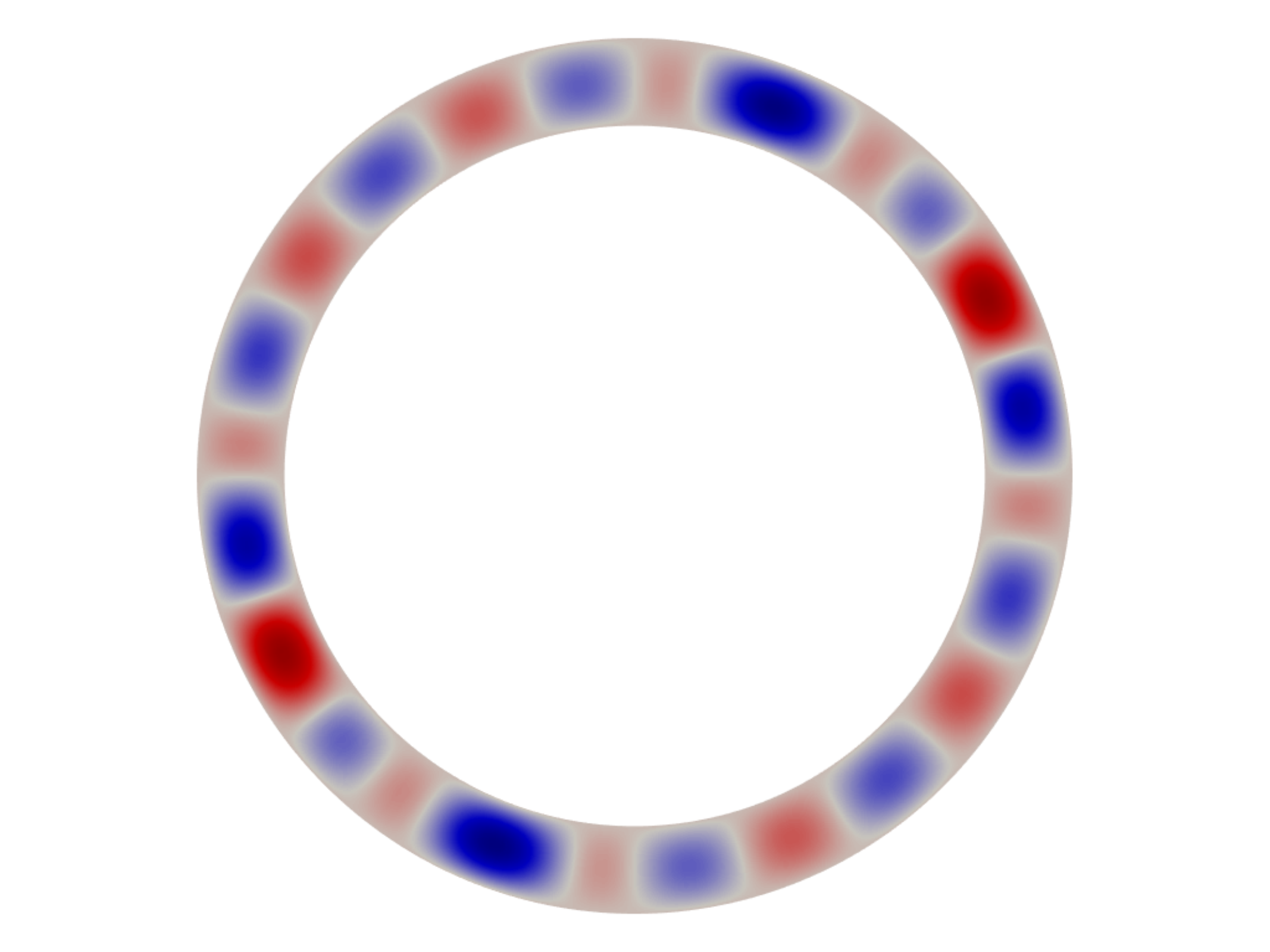}
    \end{subfigure}
\caption{2D polar slices of the synthetically generated radial velocity field with dimensionless units for a range of dominant angular wavenumbers ($\ell$ = 1, 2, 4, 6, 8, 14).}
    \label{fig:Vr_2D}
\end{figure*}

By varying the ratio of the inner and outer radii of the convection zone, we can probe the expected range of expected angular wavenumbers in different convective shells.

As has been observed in simulations of non-rotating stellar convection, the angular wavenumber $\ell$ is expected to be $\approx 1$ for a convective core \citep{Herwig2023}, and moves to larger values
as the shell gets thinner, leading to more convective plumes and smaller convective cells. We show how our velocity field model recovers this in Figure~\ref{fig:Vr_2D}, where we plot slices of the radial velocity for $\ell$ = 1, 2, 4, 6, 8 and 14. 
Once again, the radial velocity is shown in dimensionless units. We note that these can easily be converted to physical units by multiplying by a characteristic velocity scale $V_0$ and a characteristic length scale $L_0$.

\begin{figure*}
\centering
    \begin{subfigure}{0.9\linewidth}
        \includegraphics[width=\linewidth]{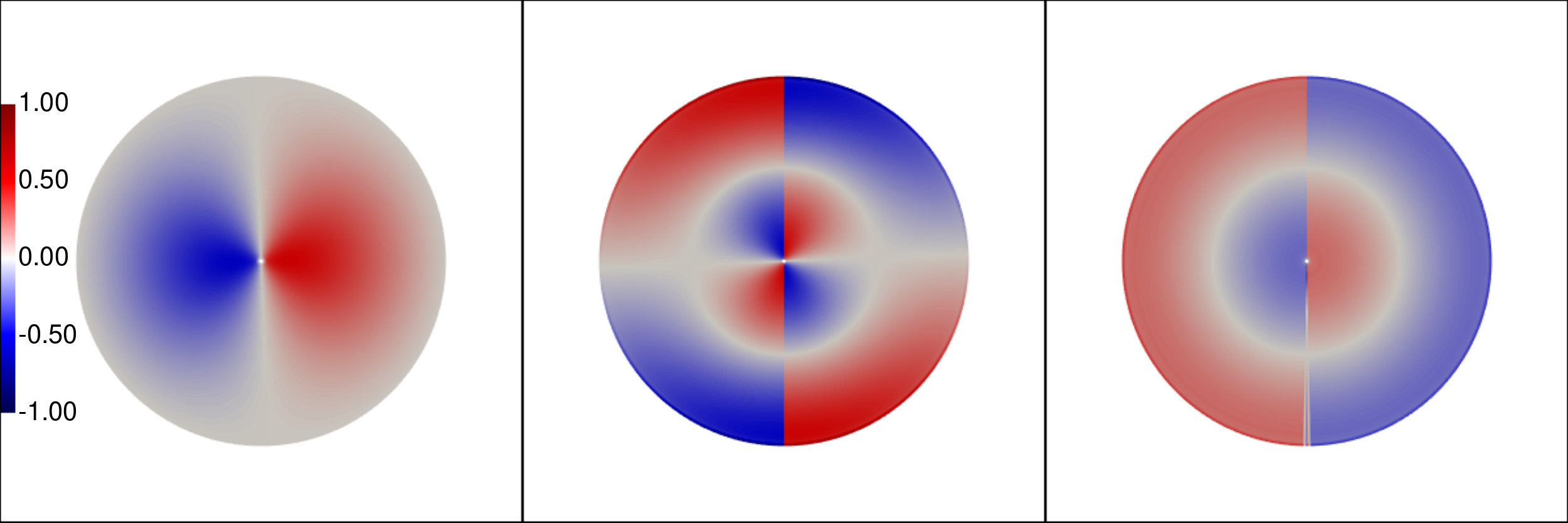}
    \end{subfigure}

    \begin{subfigure}{0.9\linewidth}
        \includegraphics[width=\linewidth]{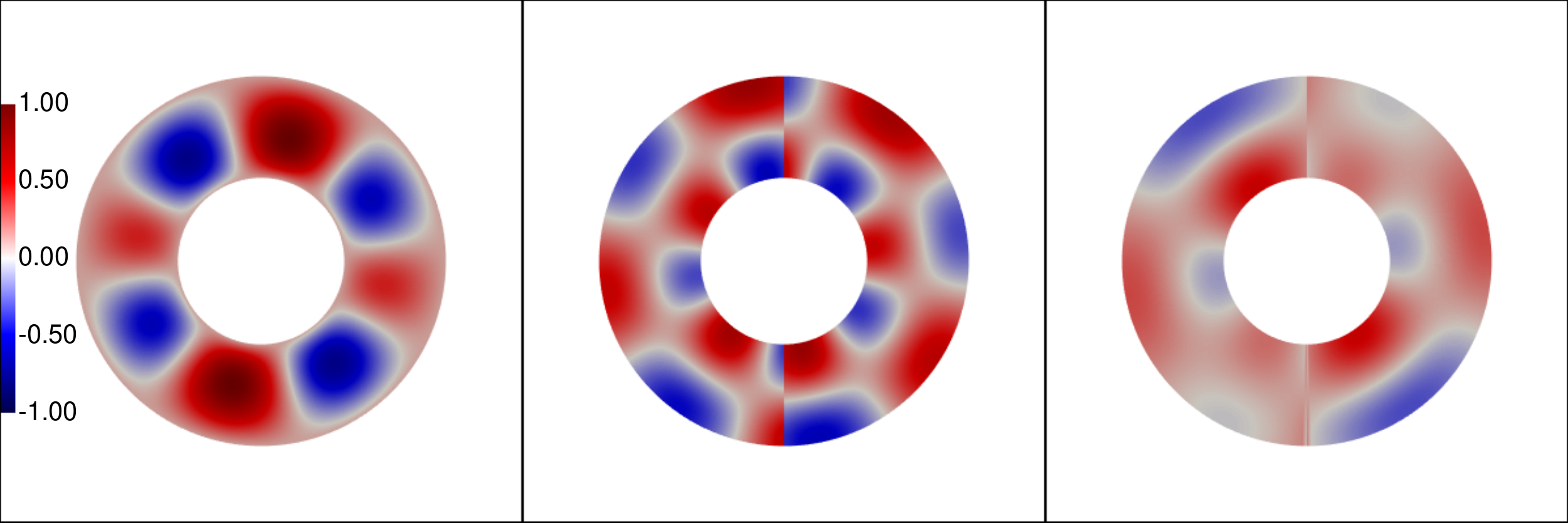}
    \end{subfigure}
\caption{2D polar slices showing all three velocity components of a dimensionless velocity field for $\ell$ = 1 and 4 (left: $v_r$, middle: $v_\theta$, right: $v_\phi$).}
    \label{fig:Vfull_2D}
\end{figure*}

For two of these cases, we also show the full velocity field in Figure~\ref{fig:Vfull_2D}, where we plot slices of all three velocity components, $v_r$, $v_\theta$ and $v_\phi$ for $\ell$ = 1 and $\ell$ = 4.
Qualitatively, we find that the velocity field resembles what we expect of convective flow, with updrafts and downdrafts in the radial direction in $v_r$, flow directed either towards or away from the poles at the convective boundaries in $v_\theta$, as the plumes 
move along the boundaries before reversing their radial directions, and similarly for $v_\phi$.
We note the strange colouration of the $v_\phi$ plots, since these are velocity slices through the poles, $v_\phi$ is directed into and out of the page, and hence changes sign across the slice.

\subsection{Stellar Models}
As the primary motivation for this work is to generate initial conditions for 3D core-collapse supernova simulations, here, we show how the velocity field can be generated for realistic stellar models.

Using \texttt{MESA} data as input, the user can manually specify the radial boundaries of the convective zone they want to calculate, or allow it to automatically determine convective zones. This is done using the "$\textrm{sch\_stable}$" column, where regions are expected to be convectively unstable using the Schwarzschild criterion. This can easily be modified for the Ledoux criterion if required. If we use data at stellar core collapse, often, the calculations become numerically unstable, and individual zones may be unrealistically regarded as convectively unstable. We have mitigated this somewhat by introducing a parameter to calculate only regions where there are at least "N = $\mathrm{conv\_len}$" subsequent convectively unstable cells. We also note that when $\ell$ is too large, the computation time becomes prohibitively long (due to the summation of all $m$ values seen in Equation~\ref{eq:psi}), which is another reason to place a lower limit on the size of the convective zones we consider.

As mentioned above, our normalised dimensionless velocity field (which has velocities from -1 to 1) can be scaled by a characteristic velocity scale $V_0$. For the case where a stellar profile is used (e.g. from the \texttt{MESA} data file), we provide two options for choosing $V_0$. Either $V_0$ is chosen automatically as the maximum MLT velocity in the convective zone of interest, or it can be provided by the user.

\begin{figure*}
\centering
    \begin{subfigure}{0.9\linewidth}
        \includegraphics[width=\linewidth]{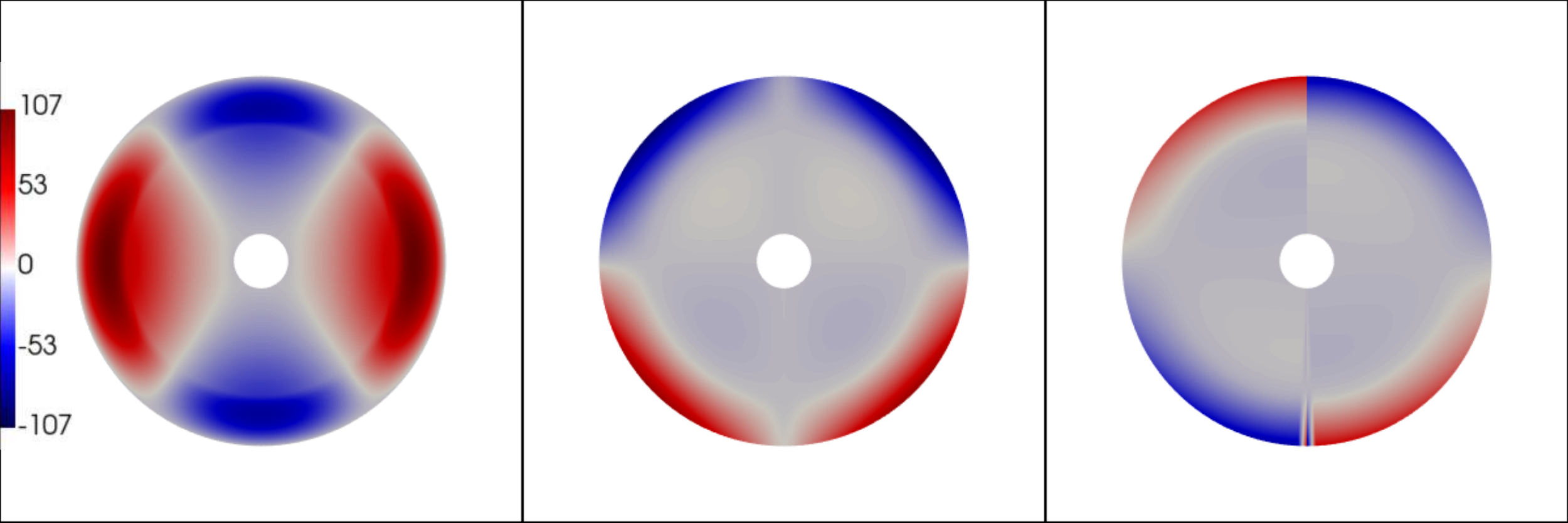}
    \end{subfigure}
    
    \begin{subfigure}{0.9\linewidth}
        \includegraphics[width=\linewidth]{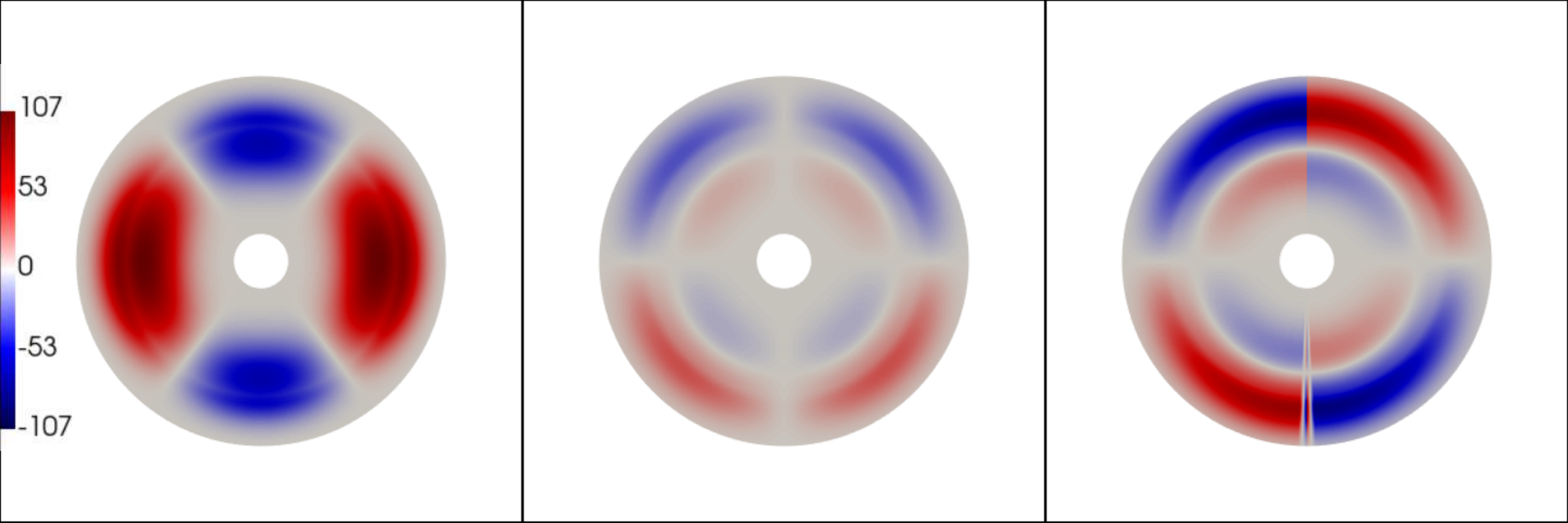}
    \end{subfigure}
    
    \begin{subfigure}{0.9\linewidth}
        \includegraphics[width=\linewidth]{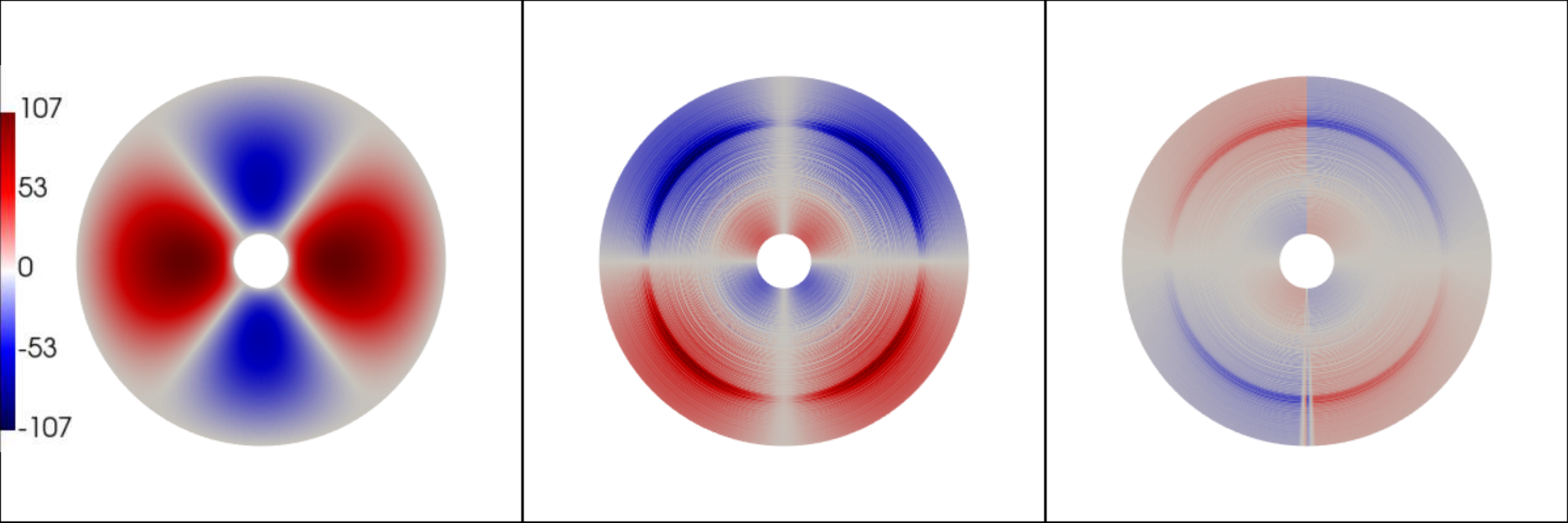}
    \end{subfigure}

\caption{Slices of $v_r$ (left), $v_\theta$ (middle) and $v_\phi$ (right) for different choices of radial basis functions, windowed sine (top), Gaussian (middle), and density weighted windowed sine (bottom) of a 16$\mathrm{M_\odot}$ \texttt{MESA} model, which is a post-merger shell, several minutes before core-collapse, with a large density stratification. Note that we do not use random phases for the complex vector spherical harmonics amplitudes here, so that the plots for the different radial basis functions can be meaningfully compared. All plots use the same colourbar, and are in units of km/s.}
    \label{fig:Radial_basis_2D}
\end{figure*}

We use stellar models from the \texttt{MESA} stellar evolution code calculated in Whitehead et.al (in-prep), and use the velocity field prescription described above to generate 3D synthetic velocity fields for some convective regions.
In Figure~\ref{fig:Radial_basis_2D}, we show slices of all three velocity field components for a convective shell of a 16$\mathrm{M_\odot}$ stellar model, at $\approx 3000$s prior to core-collapse. The velocities shown in the colourbars are in km/s, and have been scaled to the MLT velocity extracted from the \texttt{MESA} data.
The convective zone in question is a post-merger shell (after the merging of the carbon, neon and oxygen shells) with a large density stratification, where our simple estimate for $\ell$ (Equation~\ref{eq:ell_approx}) finds a quadrupolar mode for this convective shell. 
When we choose the radial basis function to be a windowed sine function (Figure~\ref{fig:Radial_basis_2D}, top row), we find an artefact where the velocity is concentrated at the outer regions of the convection zone where the density is low.
We can alleviate this artefact somewhat by choosing a Gaussian radial basis function (Figure~\ref{fig:Radial_basis_2D}, middle row), which concentrates the velocity field towards the centre of the convective shell,
or by removing the density dependence by choosing a density-weighted windowed sine function (Figure~\ref{fig:Radial_basis_2D}, bottom row), however, we are left with some "ringing" artefacts in $v_\theta$ and $v_\phi$.

While our choices in radial basis functions are physically motivated, these choices are not derived from any first-principle arguments. 
As such, we do not prescribe a specific choice of radial basis function, as this depends on the specific convective region being modelled, and the density stratification across the convective shell.
However, we note that, from our test, most thin convective shells ($\ell \gtrsim 3$) give qualitatively reasonable solutions when we choose a windowed sine function. 
Similarly, a weighted windowed sine function is a good choice for convective shells with large density stratification, as it recovers reasonable radial velocities. Although $v_\theta$ and $v_\phi$ 
show artefacts, if these fields are being used to initialise 3D simulations in practice, these artefacts should quickly diffuse away numerically once mapped to a hydrodynamic code. 

\begin{figure*}
    \centering
        \begin{subfigure}{0.49\linewidth}
            \includegraphics[width=\linewidth]{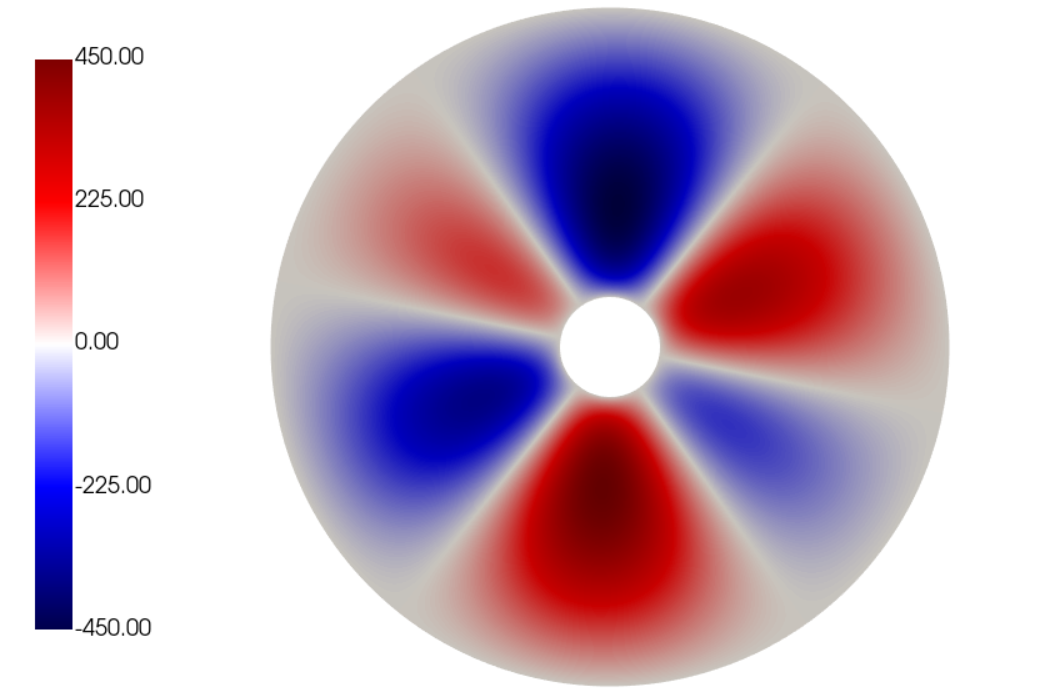}
        \end{subfigure}
    \hfil
        \begin{subfigure}{0.41\linewidth}
            \includegraphics[width=\linewidth]{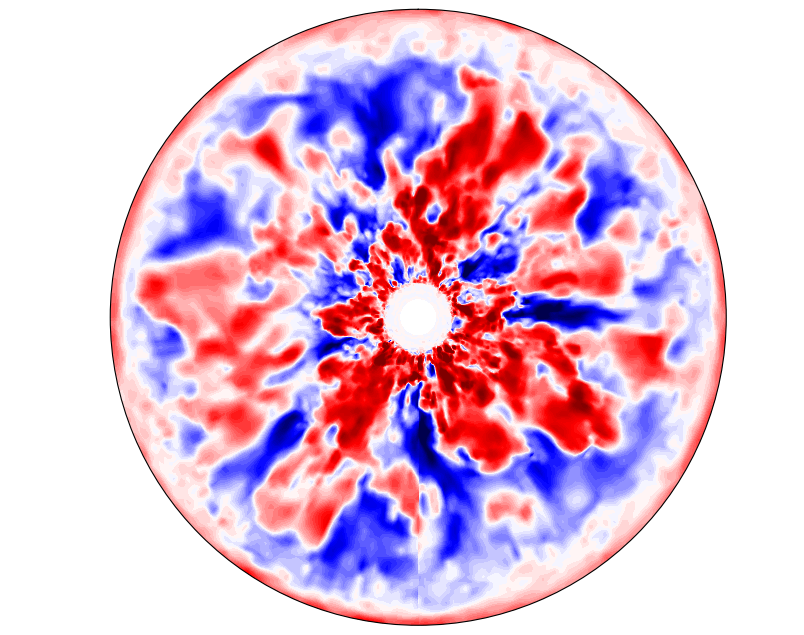}
        \end{subfigure}
    
    \caption{Slice of a 3D hydrodynamic simulation (right) and the equivalent synthetic velocity field ($\ell$ = 3) of a post-merger convective shell of a 16$\mathrm{M_\odot}$ \texttt{MESA} model (the radial basis function chosen is a weighted windowed sine function). Both plots use the same colourbar, and are in units of km/s.}
        \label{fig:m16_carbon}
\end{figure*}

\begin{figure*}
    \centering
        \begin{subfigure}{0.49\linewidth}
            \includegraphics[width=\linewidth]{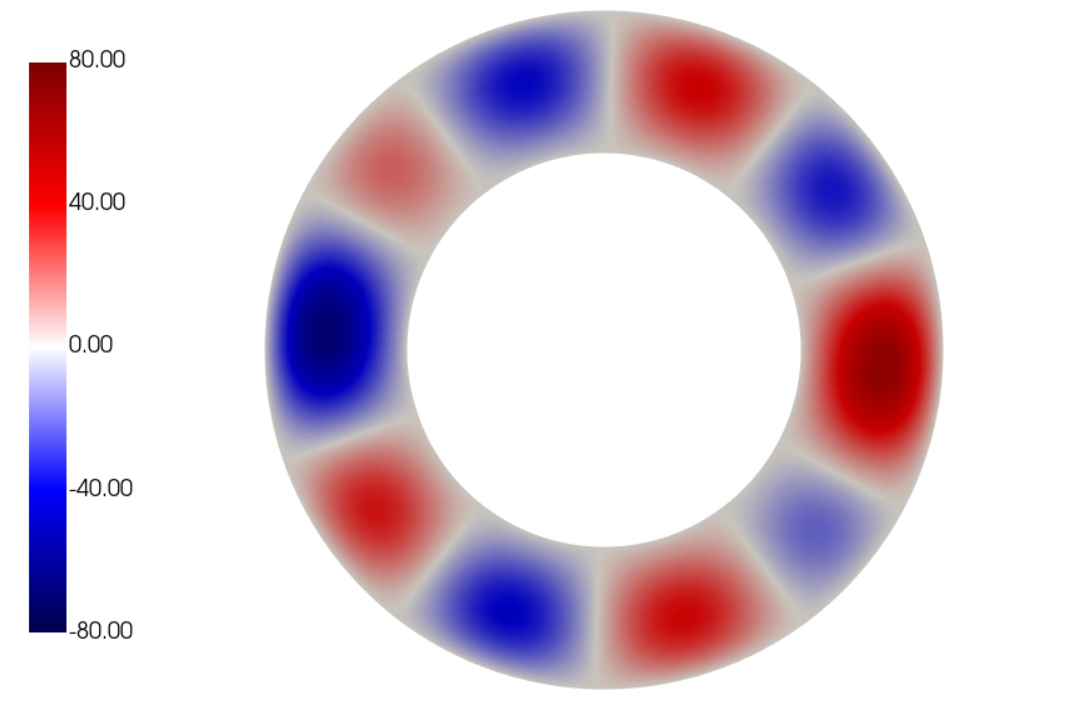}
        \end{subfigure}
    \hfil
        \begin{subfigure}{0.41\linewidth}
            \includegraphics[width=\linewidth]{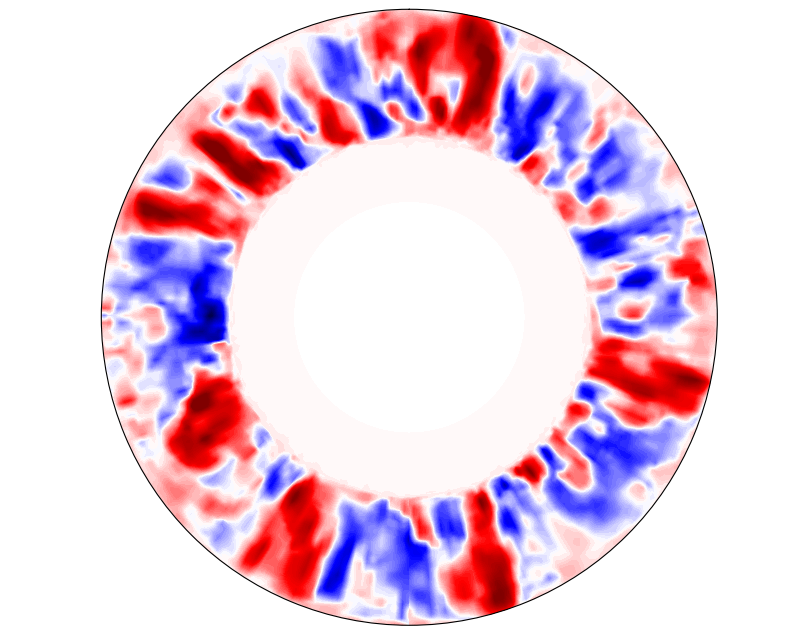}
        \end{subfigure}
    
        \caption{Slice of a 3D hydrodynamic simulation (right) and the equivalent synthetic velocity field ($\ell$ = 5) of a stable oxygen burning convective shell of a 25$\mathrm{M_\odot}$ \texttt{MESA} model, several days before core collapse (the radial basis function chosen is a weighted windowed sine function). Both plots use the same colourbar, and are in units of km/s.}
        \label{fig:m25_oxy}
\end{figure*}

As a final sanity check, we qualitatively compare two generated synthetic velocity fields to 3D hydrodynamic simulations we have run with the CoCoNuT code \citep{Varma2021, Varma2023}. We note that we have performed a spherical harmonic decomposition to determine the kinetic energy spectrum for the two simulations we present above to determine their dominant modes, which is what we use to reconstruct the synthetic velocity field. We have also chosen the characteristic velocity scale, $V_0$, of our synthetic velocity field to be the maximum velocity found in each 3D simulation.

In Figure~\ref{fig:m16_carbon}, we once again present the same post-merger convective shell of the 16$\mathrm{M_\odot}$ \texttt{MESA} model that was shown in Figure~\ref{fig:Radial_basis_2D},
compared to a slice from a 3D hydrodynamic simulation of this progenitor.
We present the radial velocity component, $v_r$, of the synthetic velocity field assuming a dominant mode of $\ell$ = 3 using a weighted windowed sine basis function, and find it to qualitatively resemble the large-scale structure of the 3D simulation of that progenitor. 
As shown previously in Figure~\ref{fig:Radial_basis_2D},  gives us an estimate of the angular wavenumber $\ell$ = 2 for this shell, Equation~(\ref{eq:ell_approx}). A spherical harmonic decomposition of the kinetic energy of our simulation shows the dominant mode to be closer to $\ell$ = 3, which we use for this comparison. As discussed previously, Equation~(\ref{eq:ell_approx}) has been shown to be compatible with linear stability analysis, so the small discrepancy between the estimate and the dominant mode of the simulation is not unexpected once non-linearities set in. 

Similarly, we compare a stable oxygen burning convective shell several days before core collapse of a 25$\mathrm{M_\odot}$ \texttt{MESA} model (from Whitehead et al. in prep) in Figure~\ref{fig:m25_oxy}, where we again find that the synthetic velocity field qualitatively resembles the 3D hydrodynamic simulation. Here, our simple estimate of the angular wavenumber $\ell$ = 5 is consistent with the spherical harmonic decomposition of the kinetic energy of the simulation, and we use this for our synthetic velocity field.

While not exact, we find that Equation~(\ref{eq:ell_approx}) is a good first-order estimate of the angular wavenumber to be used to generate synthetic velocity fields. 

\section{Conclusion}
\label{sec:conclusion}
In this work, we presented a method to generate synthetic 3D velocity fields for convective zones in stellar interiors, 
which can be used as initial conditions for 3D core-collapse supernova simulations. Using vector spherical harmonics (VSH)
and imposing physical constraints such as anelastic flow, non-radial vorticity, and zero net angular momentum, we 
construct a velocity field prescription that captures the essential features of convective flow.

Our implementation generates flow patterns by estimating a dominant angular wavenumber of the shell, $\ell$,
and superpose modes with different azimuthal numbers $m$, ensuring isotropy to produce realistic convective plume shapes. 
We have also explored various radial basis functions to model the radial dependence of the velocity field. While the choice of radial basis function depends 
on the specific stellar model and convective zone being studied, we find that windowed sine functions and density-weighted variants are particularly effective for reconstructing qualitatively realistic convective patterns.

We have validated our synthetic velocity fields by comparing them to 3D hydrodynamic simulations of convective shells 
in massive stars. These comparisons show that our method can qualitatively reproduce the large-scale flow patterns observed 
in our simulations. We note that our simple estimate for the angular wavenumber $\ell$ provides a reasonable 
first-order approximation for the dominant mode of convection, however, due to the non-linearities in full 3D simulations,
the real dominant $\ell$ modes can differ slightly. 

Overall, this method provides a computationally inexpensive alternative to full 3D stellar convection simulations, 
enabling the study of core-collapse supernovae with realistic initial conditions derived from 1D stellar evolution models.
This model can also be used by the 3D hydrodynamic community to provide seed perturbations to initialise 3D convection simulations.
An open-source \textsc{Python} implementation of the method is available on GitLab, accompanying this manuscript, which should allow 
the stellar and CCSN community to easily utilise this method to generate new initial conditions for their simulations.

As the Mach number and $\ell$-modes of the convective flow are the primary parameters that determine the impact on the 
CCSN explosion \citep{Muller2015b}, the current implementation presented should be useful for us to explore 
the important pre-supernova parameter space. However, future work will 
use 3D convection simulations to refine the radial basis functions and incorporate turbulence spectra, on top of the 
dominant convective mode, to better mimic the dynamics of buoyancy-driven convection. Additionally, we will 
extend this approach to include rotation and explore the possibility of using a similar framework to generate realistic 
initial magnetic field geometries. 

\section*{Acknowledgements}
This work used the DiRAC Memory Intensive service (Cosma8) at Durham University, managed by the Institute for Computational Cosmology on behalf of the STFC DiRAC HPC Facility (www.dirac.ac.uk). The DiRAC service at Durham was funded by BEIS, UKRI and STFC capital funding, Durham University and STFC operations grants. DiRAC is part of the UKRI Digital Research Infrastructure. BM acknowledges support from the Australian Research Council through Discovery Project DP240101786.
RH acknowledges support from the World Premier International Research Centre Initiative (WPI Initiative), MEXT, Japan, the IReNA AccelNet Network of Networks (National Science Foundation, Grant
No. OISE-1927130) and the Wolfson Foundation.

\section*{Data Availability}
The software package to solve the mathematical formulations presented to produce the synthetic velocity fields is available at \url{https://gitlab.com/vishnuvarma/Stellar_velocity_field}. The package also includes routines along with the \texttt{MESA} profile data required to reproduce the plots in this manuscript. The GitLab page includes some documentation to help the user get started. The simulated data presented in this article will be shared on reasonable request to the authors, subject to considerations of intellectual property law.



\bibliographystyle{mnras}
\bibliography{paper} 
\bsp	
\label{lastpage}
\end{document}